\documentclass[12pt]{article}
\usepackage{graphicx}
\usepackage{amssymb,amsmath}
\usepackage{epsfig}
\def\half{\frac{1}{2}}

\def\C{\mathbb{C}}

\def\<{\langle}
\def\>{\rangle}

\def\P{\mathbb{P}}

\title{Crystals and intersecting branes}
\author{Daniel L. Jafferis {\footnote{E-mail: jafferis@string.harvard.edu}}\\
Jefferson Physical Laboratory\\Harvard University\\Cambridge, MA 02138}
\date{July 5, 2006}
\begin{document}

\maketitle

\abstract{We show that the index of BPS bound states of D4, D2 and
D0 branes in IIA theory compactified on a toric Calabi Yau are
encoded in the combinatoric counting of restricted three
dimensional partitions. Using the torus symmetry, we demonstrate
that the Euler character of the moduli space of bound states
localizes to the number of invariant configurations that can be
obtained by gluing D0 bound states in the $\C^3$ vertex along the
D2 brane wrapped $\P^1$ legs of the toric diagram. We obtain a
geometric realization of these configurations as a crystal
associated to the extra bound states of D0 branes at the singular
points of a single D4 brane wrapping a high degree equivariant
surface that carries the total D4 charge. We reproduce some known
examples of the partition function computed in the opposite regime
where D0 and D2 charge are dissolved into D4 flux, as well as
significantly generalize these results. The crystal
representation of the BPS bound states provides a direct
realization of the OSV relation to the square of the topological
string partition function, which in toric Calabi Yau is also
described by a theory of three dimensional partitions.}

\begin{section}{Introduction}

The recent conjecture of Ooguri, Strominger, and Vafa \cite{OSV} that the indexed entropy of ${\cal N}=2$ black
holes, in a mixed ensemble, factorizes, perturbatively to all orders for large charges, into the square of the
topological string, together with subsequent checks for the case of IIA compactified on local Calabi Yau in
\cite{Vafa} \cite{AOSV} \cite{AJS} has generated much renewed
interest in the structure of the partition functions of twisted
${\cal N}=4$ Yang-Mills living on D4 branes. In this work, we will
discover a situation in which this theory localizes to a discrete
sum over equivariant configurations, associated to a new type of
melting crystals, related to those of \cite{topcrystal} which
describe the topological A-model in the same geometry. Taking the
limit of large D4 brane charge automatically reproduces the square
of the A-model crystals, at the attractor values of the Kahler
moduli.

Consider D4 branes wrapping various 4-cycles of a toric
Calabi-Yau, with chemical potentials for D2 and D0 branes turned
on. The partition function, studied in \cite{VW}, is the generating function of the
number (really index) of BPS bound states, which is given by the
Euler characteristic of their moduli space. This can be computed
in the $\mathcal{N} =4$ worldvolume $U(N_i)$ Yang Mills on the
4-cycles wrapped by $N_i$ D4 branes, with some bifundamental
couplings along the intersections. We will find that, at least
locally in a single vertex geometry, the number of bound states is
captured by the combinatoric counting of three dimensional
partitions subject to a certain truncation.

From the perspective of the remaining four dimensional spacetime,
these bound states appear as BPS black holes, preserving half of
the ${\cal N}=2$ supersymmetry, possessing a classical horizon with nonzero
area in the regime of large charges. The Kahler moduli live in
vector multiplets in IIA theory, hence they are driven to their
attractor values at the horizon, which are determined entirely by
the brane charges, via the quantum corrected prepotential. The Legendre transformation leading to a mixed
ensemble, with chemical potentials $\phi^0$ and $\phi^a$ for the
D0 and D2 charges, results in attractor values that are
independent of the prepotential,
\begin{equation} g_{top} = \frac{4 \pi^2}{\phi^0 - i \pi p^0}, \
t_a = g_{top} (p^a + \frac{i}{\pi} \phi^a ), \end{equation} as
shown in \cite{OSV}. The BPS partition function in this ensemble
factorizes in the large charge limit according to the OSV
relation, \begin{equation} Z_{BPS} (\phi^0, \phi^a; p^0, p^a) =
|\psi_{top} (g_{top}, t_a)|^2 , \end{equation} to be interpreted
as a perturbative expansion in the topological string coupling,
$g_{top} \sim 1 / \phi^0$. We will be concerned with the case of
vanishing D6 charge, $p^0=0$.

The torus symmetry of a toric Calabi Yau induces an action on the
moduli space of branes, enabling us to apply the equivariant
localization theorem that the Euler character of the full moduli
space is the same as that of the space of invariant
configurations. Therefore we can assume that the D0 branes wrap
the fixed points, and the D2 branes wrap fixed curves, which are
the $\P^1$ legs of the toric diagram. Imagine cutting the Calabi
Yau along the toric legs, so that it looks like copies of the
local $\C^3$ vertex glued along $\P^1$'s over which the manifold
is fibered as ${\cal O}(-p) \oplus {\cal O} (p-2)$ for an integer
$p$ sometimes called the framing. Then the bound states of D2
branes can be recovered by matching equivalent configurations at
the cut, and D0 branes are localized to a single vertex. This is
very reminiscent of the toric localization \cite{AKMV} of the
topological string amplitudes, particularly as computed by the
quantum foam D6 brane theory \cite{qfoam}.

The generic points in the moduli space of very ample D4 branes can be reached by turning on VEVs for the ${\cal N}
= 4$ adjoint field describing deformation in the normal direction inside the Calabi Yau, leading to a
configuration of a single D4 brane wrapping a complicated smooth surface, ${\cal C} \subset X$, whose homology is
fixed by the brane charges. In this phase, the bound states that were described in terms of D2 and D0 branes
dissolved into flux of the $\prod U(N_i)$ quiver theory of intersecting D4 branes, become $U(1)$ flux
configurations associated to the large number of new divisors in the deformed surface. The "D2" bound states
wrapping curves in $H_2 ({\cal C}) > H_2 (X)$ which are homologically trivial in the ambient 6 dimensional
geometry carry no D2 charge, and replace the D0 flux bound states present in the $\prod U(N_i)$ phase.

It is easy to see that such smooth D4 surfaces beak the $(\C^*)^3$
symmetry of the toric Calabi Yau. The crystal construction arises
when we consider nilpotent VEVs for the adjoint field, which
allows the use of the underlying toric symmetries to solve the
theory. This corresponds to the limit in which the new divisors in
${\cal C}$ shrink to zero size, giving rise to the nontrivial D0
bound states at the singular points that will be counted by the
crystal.

The bound states of D2 and D0 branes to a single stack of $N$ D4 branes wrapping a noncompact cycle in a toric
Calabi Yau was studied in \cite{AOSV}, where the theory was reduced to q-deformed Yang Mills in two dimensions.
The D2 bound states wrapping a $\P^1$ are parameterized by irreducible representations of $U(N)$, which are
associated to the holonomy basis of the Hilbert space of the two dimensional topological gauge theory. These
configurations can be readily understood in the Higgs phase of the D4 theory, describing the generic point in the
moduli space of normal deformations where there is a single D4 brane wrapping a complicated surface. In that case,
the D4 worldvolume is a N-sheeted fibration over the $\P^1$, and hence, locally, the transverse motion of $k$
bound D2 branes is described by $k$ points in the fiber, with simply looks like $N$ copies of $\C$. Note that for
the purpose of this discussion we can ignore the global obstruction to deforming a curve inside the D4
worldvolume. This is because the partition function separates into propagator and cap contributions, and the D2
bound states can be equivalently analyzed in the propagator geometry, over which the bundles are trivial. Counting
the number of ways of distributing $k$ objects among $N$ choices gives the relevant $U(N)$ representations,
associated to the moduli space
\begin{equation} \frac{\textit{Sym}^k (N \C) }{S_N}.
\end{equation} We will see this feature more clearly by choosing a toric invariant D4 worldvolume in which to
perform the analysis, where these bound states will be described by certain torus equivariant ideals.

Furthermore, there is an interesting and suggestive structure to
the generating function of bound states in the vertex geometry of
a stack of $N$ D4 branes with only D0 branes. In the q-deformed
Yang Mills language, this is the vacuum cap amplitude, that is,
the amplitude with trivial holonomy, which has been calculated in
\cite{AOSV} to be
\begin{equation} \eta(q)^{-N \chi} S^{(N)}_{\cdot \cdot} =
\frac{1}{\eta^{N}} \sum_{\sigma \in S_N} (-)^\sigma q^{-\rho_N
\cdot \sigma(\rho_N)} = q^{-N/24} \prod_{n>0} \frac{1}{(1-q^n)^N}
\prod_{1 \leq i<j \leq N} [j-i]_q . \end{equation} The BPS states
in which the D0 charge is dissolved into $U(N)$ flux are captured
by the q-deformed Yang Mills amplitude, while the $\eta$ function
counts one state for each $k$-tuple of point-like D0 branes bound
to one of $N$ identical D4 branes, noting that the effective Euler
characteristic of the cap is $\chi=1$. This amplitude can be
expressed as
\begin{equation} \prod_{n>0} \frac{1}{(1-q^n)^N} \prod_{j=1}^{N-1} (q^{j/2}-q^{-j/2})^{N-j} = \prod_{j=1}^{N}
\frac{1}{(1-q^j)^j} \prod_{j>N} \frac{1}{(1-q^j)^N},
\end{equation} associated to bound states of $k$-tuples of D0 branes with $k$ ground states, truncated at $N$.
This function nicely interpolates between the $1/\eta$ for a single D4 brane, and asymptotes to the McMahon
function, which counts bound states to a D6 brane. The truncation has a transparent interpretation in terms of the
crystal we will now proceed to describe.

Consider the generic situation of $N,M$, and $K$ D4 branes
wrapping the three intersecting 4-cycles in the local $\C^3$
vertex, with bound D2 branes in the legs, and $n$ D0 branes at the
vertex. We will first find that the equivariant BPS bound states
of $p$ D2 branes wrapping the intersection curve of $N$ and $M$ D4
branes are partitions of $p$ that do not contain the point
$(M+1,N+1)$. In the special case $M=0$, this reduces to Young
tableaux with $N$ rows, which are related to the $U(N)$
representations that span the q-deformed Yang Mills Hilbert space.
Such bound states have D2 charge $p$, and induce some D0 charge
which depends on the local bundle over the wrapped curve via the
intersection form. Fixing such Young diagrams, $R,Q$, and $P$, as
the asymptotic D2 brane configuration in the vertex geometry, we
will show that each contribution to the index of BPS states
carrying an additional $n$ units of D0 brane charge can be
associated to a three dimensional partition of $n$ in the
truncated region of the octant obtained by deleting all the points
beyond $(N+1,M+1,K+1)$, with asymptotic behavior $R,Q,P$, as shown
in figure 4.

The organization of this paper is as follows: In section 2, we review the theory of BPS D4 brane bound states,
particularly in the local toric geometries studied in \cite{Vafa} \cite{AOSV} and \cite{AJS}, and the OSV relation
to topological strings \cite{OSV}. In section 3, we will examine the worldvolume theory on these branes in the
Higgs branch. Giving a nilpotent VEV to the adjoint field, which controls normal deformations inside the Calabi
Yau, reduces the partition function to a sum over toric invariant ideal sheaves of certain singular surfaces. In
section 4, we study the bound states with D2 branes in the toric "leg". In section 5, we compute the index of D0
bound states in the vertex geometry. In section 6, we see that the OSV factorization is manifestly verified in all
cases where these methods apply. Finally, in section 7, we make some concluding remarks.

\end{section}

\begin{section}{BPS bound states of the D4/D2/D0 system}

The remarkable conjecture of OSV \cite{OSV} relates the large
charge asymptotics of the indexed degeneracy of BPS bound states
making up supersymmetric black holes in four dimensions with the
Wigner function associated to the topological string wavefunction.
Consider compactifying IIA theory on a Calabi Yau manifold, $X$,
which gives rise to an effective ${\cal N}=2$ supergravity in the
remaining four dimensions, with vector multiplets associated to
the Kahler moduli. The microstates of $\half$ BPS black holes in
this theory are bound states of D6/D4/D2/D0 branes wrapping
holomorphic cycles in $X$.

The effective theory in four dimensions has $U(1)$ gauge fields obtained by integrating the RR 3-form, $C_3$, over
 2-cycles, $[D^a] \in H_2(X)$ for $a = 1, \dots h^{1,1} (X)$, in the Calabi Yau. The D2 branes wrapping these 2-cycles
 and D4 branes on the dual 4-cycles, $[\check{D}_a]$, carry electric and magnetic charges respectively under these
 $U(1)$'s. There is one further $U(1)$ vector field, arising from the RR 1-form in ten dimensions, under which the
 D0/D6 branes are electrically (magnetically) charged.

We will investigate this partition function at $0$ D6 charge in the local context of toric Calabi Yau, where the
theory on the branes is given by twisted ${\cal N} = 4$ Yang Mills on each stack of D4 branes, coupled along their
intersections. The action of the $U(N)$ worldvolume theory of $N$ D4 branes is BRST exact, and turning on the
chemical potential for bound D2 and D0 branes inserts the closed (but not exact) operators, \begin{equation}
\frac{\phi_0}{8 \pi^2} \int F \wedge F + \frac{\phi_a}{2 \pi} \int F \wedge k_a , \end{equation} into the action.
These topological terms count the induced D2 and D0 charge of a given gauge connection.

This Vafa-Witten theory \cite{VW} computes the indexed degeneracy of BPS bound states with partition function equal to
\begin{equation} Z_{YM}({Q_4}^a, \varphi^a, \varphi^{0}) = \sum_{Q_{2 \, a},Q_0} \ \Omega(Q_4^a,Q_{2\, a}, Q_0) \ \
\exp \left( - Q_0 \varphi^0 - Q_{2 \, a} \varphi^{a} \right),
\end{equation} in the mixed ensemble of fixed D4 (and vanishing D6)
charges.

A collection of D4 branes in the OSV mixed ensemble will be well described by semiclassical ten dimensional
supergravity as a black hole with large horizon area when the branes live in a homology class deep inside the
Kahler cone \cite{MSW}. Given D4 branes wrapping such a very ample divisor, $[D]$, one can show that the attractor
values of the moduli are large and positive, \begin{equation} \mathfrak{Re} (t^a) >>0, \end{equation} as well as the
leading classical contribution to the black hole entropy, \begin{equation} C_{abc} t^a t^b t^c > 0,
\end{equation} since the intersections numbers of $[D]$ with 2-cycles are positive. It is in this regime where we
expect the brane partition function to factorize into the square
of the topological string wavefunction. Moreover, these are
precisely the conditions on the 4-cycle, with homology $[D]$ that
is wrapped by the D4 branes, to have holomorphic deformations. The
existence of such global sections of the $(2,0)$ bundle over the
4-cycle in the compact case would allow us to turn on a
topologically exact mass deformation and solve the theory
following \cite{VW}. For branes wrapping noncompact 4-cycles, the
mass deformation results in the reduction to a theory on the two
dimensional compact divisors.

This two dimensional theory was shown to be the q-deformed $U(N)$
Yang Mills on the divisor of genus $g$ in \cite{Vafa},
\cite{AOSV}, who then solved the theory using techniques of
topological gauge theory in two dimensions, obtaining the result
\begin{equation} Z_{YM} = \sum_{{\cal R}} ( \dim_q {\cal R}
)^{2-2g} q^{\frac{p}{2} C_2({\cal R})} e^{i \theta |{\cal R}|},
\end{equation} where ${\cal R}$ is an irreducible $U(N)$
representation. The more general situation of D4 branes
intersecting in the fiber over the intersection point of two
curves was solved in the toric context by \cite{AJS}, where the
q-deformed Yang Mills theories found on each curve were coupled by
operators inserted at the intersection point, obtained from
integrating out bifundamentals. In this work, we will check the
results of the crystal partition function against these examples,
as well as generalizing to configurations of D4 branes
intersecting along the base.

The BPS partition function in the large charge limit was shown to
factorize, confirming the generalization of the structure $Z \sim
|\Psi_{top}|^2$ to the noncompact case with an extra sum over
chiral blocks parameterized by $SU(\infty)$ Young tableaux, $P$,
$P'$ controlling the noncompact moduli. In particular, for the
case of $N$ D4 branes in  the Calabi Yau ${\cal O}(-p) \oplus
{\cal O}(p-2) \rightarrow \P^1$ studied in \cite{AOSV}, the large
$N$ 't Hooft limit gives
\begin{equation}Z=\sum_{\ell\in\mathbb{Z}}\sum_{P,P'}Z^{qYM,+}_{P,P'}(t+pg_s\ell)Z^{qYM,-}_{P,P'}(\bar{t}-pg_s\ell),
\label{tsfact}\end{equation}
where the topological string amplitudes are \begin{equation}\begin{split}Z^{qYM,+}_{P,P'}(t)&=q^{(\kappa_P+\kappa_{P'})/2}
e^{-\frac{t(|P|+|P'|)}{p-2}}\\
&\qquad\times\sum_R q^{\frac{p-2}{2}\kappa_R}e^{-t|R|}W_{PR}(q)W_{P^{\prime t}R}(q)\\
Z^{qYM,-}_{P,P'}(\bar{t})&=(-1)^{|P|+|P'|}Z^{qYM,+}_{P^t,P^{\prime
t}}(\bar{t}), \end{split}\end{equation} and
\begin{equation}g_s=\frac{4 \pi^2}{\phi_0} \qquad t=\frac{p-2}{2}
\frac{4 \pi^2}{\phi_0} N-2 \pi i \frac{\phi_2}{\phi_0},
\end{equation} are the attractor values of the moduli.

Recall that in toric Calabi Yau all compact 4-cycles have positive
curvature, and thus negative self-intersection number. Branes
wrapping these rigid cycles will contribute negative real part to
the attractor moduli. It is possible, however, that even in this
rigid case a modified local analysis is still possible, in which
an additional constraint for each wrapped compact 4-cycle is
introduced on the sum of local bundles over its toric divisors.
For example the twisted ${\cal N}=4$ Yang Mills on the rigid
divisor $\P^2$ in the Calabi Yau ${\cal O}(-3) \rightarrow \P^2$
is solved in \cite{VW}, \cite{Klyachko}, and
\cite{Klyachko2}.\footnote{We thank C. Vafa and M. Aganagic for
discussions that led to the statements here.} These partition
functions do not factorize, and correspond to the regime in which
nonperturbative baby universe effects are not suppressed
\cite{DGOV} \cite{AJS}.

More applicably in the context of OSV, some of the branes might
wrap rigid surfaces, as long as the total charge is the in Kahler
cone. In this case one can still expect a factorization into the
topological and anti-topological string, and a simplification of
the Yang Mills theory on the compact rigid surfaces may occur due
to the intersection with sufficiently large numbers of branes
wrapping very ample divisors.

Another interesting feature of the D4 brane partition functions is
their behavior under s-duality of the twisted theory in four
dimensions. This is not manifest in the crystal formalism, since
in our situation of noncompact D4 branes, the boundary conditions
at infinity in the noncompact directions transform in a
complicated way. It would be interesting to understand more
completely how the duality is realized in our picture, although
one aspect is already clear. Namely, we are really counting the
vector bundles associated to the \textit{dual} flux of the
topological Yang-Mills theory. This explains why our expansions
appear most naturally in terms of $q= e^{-g_s}=e^{-\frac{4
\pi^2}{\phi^0}}$ rather then $\tilde{q} = e^{-\phi^0}$. We will
also use the variable $\theta_a = \frac{g_s}{2 \pi} \phi^a$.

The boundary conditions that are naturally associated to the
crystal description are that no bound D2 branes should reach
$\infty$ in the noncompact D4 worldvolume. This turns out to be
the same condition used in the derivation of q-deformed Yang Mills
\cite{Vafa} \cite{AOSV}, but is the opposite of the standard
"electric" convention from the four dimensional worldvolume point of view,
which would require the holonomies of the gauge field to vanish
asymptotically. These are exchanged by the action of s-duality.
For example, the instantons contributing to the $U(1)$ Yang Mills
amplitudes on ${\cal O} (-p) \rightarrow \P^1$, expressed in terms
of the gauge fields, are given by the lift of line bundles on the
$\P^1$, with noncompact toric divisors lying in the fiber over the
fixed points. The dual configuration has the $\P^1$ itself as
toric divisor, and it is this that we shall count.

\subsection{Turning on a mass deformation for the adjoints}

The deformations of the wrapped surface are encoding in the adjoint fields of the twisted ${\cal N} = 4 \ $ $U(1)$ gauge
theory living on the D4 brane. The torus symmetry of the underlying geometry can be used to localize the D4 brane
to the equivariant configuration, by the usual method of turning on a $Q$-trivial mass deformation,
\begin{equation} W = m U V + \omega T^2, \label{massdef} \end{equation} parameterized by the choice of $\omega$, a
global section of the $(2,0)$ bundle over ${\cal C}$, where $U$ and $T$ are scalar adjoint fields associated to
motion in spacetime, while $V$ governs deformation normal to ${\cal C}$ inside the Calabi-Yau. The superpotential,
$W$, must live in a $(2,0)$ bundle over ${\cal C}$.

In our case, there is an important new feature to this story, due
to the compactification of the normal direction in the Calabi-Yau.
The adjoint field, $V$, will take values in some compact space,
which, in general, would greatly complicate the analysis. As
explained in \cite{AJS} this issue can be avoided when the D4
branes wrap noncompact cycles, since one is free to choose
boundary conditions at infinity corresponding to a particular
surface in the given homology class. Changes of these boundary
conditions are determined by non-normalizable modes in the four
dimensional theory on the branes, and we should not sum over them,
at least in the local case. It is sensible to choose this divisor
to be torus invariant, for consistency with the toric symmetry of
the local geometry (in fact, it is only in the equivariant sense
that we can even appropriately distinguish various 4-cycles which
would otherwise appear homologous in the noncompact Calabi-Yau).

It is gratifying to understand this point in greater detail, for
concreteness in the example of local $\P^1 \times \P^1$. Let us
consider a single D4 brane wrapped on the 4-cycle, ${\cal C}$,
given by ${\cal O} (-2) \rightarrow \P^1$, whose normal direction
is a compact $\P^1$. This seems to imply that the field, $V$,
should be $\P^1$ valued, however this is modified by the fact that
our 4-cycle, ${\cal C}$, is nontrivially fibred over this $\P^1$.
The local coordinate normal to the brane is thus a section of the
${\cal O} (-2)$ bundle, rather then the trivial bundle. Globally,
this implies that equation \eqref{massdef} will have two saddle
points where this section vanishes. In order to preserve the torus
symmetry, we must choose an equivariant section of the ${\cal O}
(-2)$ bundle, which implies that the mass deformation localizes
the D4 brane to exactly the two invariant surfaces with the
homology class of $[{\cal C}]$, namely those sitting above the
north and south pole of the other $\P^1$, exactly as expected.

\end{section}

\begin{section}{Localization in the Higgs branch of the theory of intersecting branes}

The worldvolume theory of D4 branes can be localized, due to the topological nature of the partition function, to
a computation of the Euler characteristic of the moduli space of anti-self dual instanton solutions. This is
because the action of the twisted theory is exact in the cohomology of the twisted supercharge, $Q$, and the
chemical potentials for D0 and D2 branes couple to the instanton charge, $\int F \wedge F$, and $\int_{[D]_a} F$
respectively, which are topological invariants. It can be shown that these anti-self dual connections exist
uniquely for each stable $U(N_i)$ vector bundle (or certain sheaves in a more general, singular context).

It is also necessary to take into account the contribution of
additional bound D0 branes, which are given by sheaves with
singularities at points where the D0 branes are located. This
gives rise to a universal contribution of $\eta^{- \chi}$ of the
eta function to the power of minus the Euler character
\cite{Nakajima}. The eta function appears as the generating
function of Young tableaux associated to the Hilbert scheme of
points in two dimensions. Such effects will be automatically
included in our case, where all of the bound states will have to
be described exclusively in the sheaf language, since the D4
branes will wrap singular surfaces.

Any collection of intersecting D4 branes whose total homology class is very ample can be deformed in the normal
directions, generically producing a single D4 brane wrapping a complicated surface. Thus the quiver of $U(N_i)$
twisted $\mathcal{N}=4$ Yang Mills theories with bifundamental interactions can be replaced by the $U(1)$ gauge
theory on this high degree 2-fold, ${\cal C}$, which computes the same partition function. In both cases the
adjoint field associated to motion of the brane in the Calabi-Yau directions will generally be compactified in a
highly nontrivial manner.

For example, consider wrapping $N$ D4 branes on the cycle ${\cal O}(-3) \rightarrow \P^1$ in the Calabi Yau ${\cal
O}(-3) \oplus {\cal O}(1) \rightarrow \P^1$, and $M$ on the rigid ${\cal O}(1) \rightarrow \P^1$ cycle. Then the
attractor value of the Kahler modulus, $\mathfrak{Re} \ t_{\P^1} = \frac{g_s}{2} (N - 3M)$, will be positive for
$N
> 3M$, hence we should be able to write down a smooth surface with these charges. The projective coordinates are
given by $(z,v;x,y) \sim (\lambda z, \lambda v; \lambda x, \lambda^{-3} y)$, where $z=v=0$ is deleted. The
deformed 4-cycle is described an equation of the form \begin{equation} x^N y^M + a x^{N-3} y^{M-1} + \dots + b
z^{N-3M} +  \dots + c v^{N-3M} = 0, \end{equation} where each term has degree $N-3M>0$.

Therefore we really want to calculate the generating function for the Euler characters of the moduli space of
ideal sheaves (ie. rank 1 stable sheaves) on ${\cal C}$. To do this, we will extensively use the toric geometry of
the underlying Calabi Yau manifold. For such techniques to be applicable, the surface, ${\cal C}$, must be chosen
carefully to be invariant itself. For a high degree surface with positive self-intersection of this type, the only
possibility is a highly singular realization as a "thick" subscheme living on the original intersecting toric
divisors. This is achieved in the worldvolume gauge theory by giving a nilpotent VEV to the adjoint field $V$ that
controls the normal deformations inside the Calabi Yau.

The ideal sheaves on such an affine space are equivalent to ideals of the algebra of functions, and can be counted
in the equivariant setting as described below. By cutting the subscheme ${\cal C}$ into pieces, we can obtain the
full answer by gluing vertices via propagators, analogously to \cite{qfoam}. Note that toric localization is
crucial for this procedure to work, with D0 branes localizing to the vertices, and D2 branes to the legs. The
vertex amplitudes, associated to bound states of D0 branes with the D4 and D2 branes, in the chemical potential,
will turn out to be nicely encoded as the statistical partition function of a cubic crystal in a restricted domain.

\subsection{Solving the twisted ${\cal N}=4$ Yang Mills theory}

The key point is that we have replaced the complicated system of
intersecting stacks of $N_i$ D4 branes described by
bifundamental-coupled $U(N_i)$ Yang Mills theories by a single D4
brane wrapping a nonreduced subscheme of $X$. The $U(1)$ theory on
that brane can be solved using toric localization, and the well
known fact that the Vafa-Witten theory computes the generating
function of stable bundles. The partition function is given by
\begin{equation} Z_{YM} = \sum_{n \in H^0, [A] \in H^2} q^n e^{i \sum_a \theta_a A_a} \chi \left( {\cal M}_{n,[A]}
 \right), \end{equation} where ${\cal M}_{n,[A]}$ is the moduli space of coherent sheaves, $n$ is the instanton number
(equivalently, the D0 charge), $[A]$ is the D2 charge, and $a$
indexes $H_2$. The effective number of bound states with given
charges is given by the Euler characteristic of the instanton
moduli space. The brane charges are given exactly by the
coefficients of the exponential Chern character, which is, in our
situation,  \begin{equation} \begin{split} q_{D0} &= 2 \it{ch}_3 =
c_3-c_1 c_2 +\frac{1}{3} c_1^3  \\ q_{D2} &= \it{ch}_2 =
-c_2+\frac{1}{2} c_1^2 \\ q_{D4} &= \it{ch}_1 = [{\cal C}] \\
q_{D6} &= 0 ,
\end{split} \end{equation} where we are ignoring the gravitation contribution from the first Pontryagin class of the Calabi
Yau, which gives an overall contribution independent of the D2/D0
bound state, and is in fact somewhat ambiguous in the noncompact
setting.

We want to determine the contribution of D0 branes bound to the
vertex in the toric geometry the looks locally like $\C^3$. In
order to be able to apply the machinery of equivariant
localization, we must choose a torus invariant representative of
the homology class $[{\cal C}]$. This is impossible for smooth
surfaces, thus we are forced to consider the singular, nonreduced
subscheme described by the polynomial equation $x^N y^M z^K = 0$,
where $N,M,K$ are the number of D4 branes wrapping the equivariant
homology classes associated to the three planes in $\C^3$. Because
the worldvolume is so singular, we need to be careful when
determining the moduli space of line bundles. Looking at the ring
of functions on the surface, it is clear that it will approach the
algebra \begin{equation}{\cal A} = \frac{\mathbb{C}[x,y,z]}{(x^N
y^M z^K)}\label{surf}\end{equation} as we deform a smooth surface
to the singular limit. This corresponds to a non-reduced subscheme
of $\mathbb{C}^3$, since the polynomial $x^N y^M z^K$ does not
have separated roots, which translates the fact that the adjoint
fields of the gauge theory have nilpotent, non-diagonalizable
VEVs.

Given any very ample line bundle on the affine scheme ${\cal C}$,
we can associate to it in the usual way an ideal of ${\cal A}$ by
considering its sections. For a general line bundle, ${\cal E}$,
corresponding to asymptotic representations with negative row
lengths along a given curve, we first take the tensor product with
a line bundle, ${\cal L}_n$, over the D4 worldvolume of
sufficiently positive first Chern class, $n$, to obtain a very
ample bundle. The bundle ${\cal L}_n$ must descend from a bundle,
${\cal O} (n)$, on the physical 4-cycle, namely, its divisor is
simply the multiply wrapped curve in question. The instanton
number can easily be calculated to be \begin{equation} \it{ch}_2
({\cal E} \otimes {\cal L}_n) = \it{ch}_2 ({\cal E}) + c_1 ( {\cal
L}_n) c_1 ({\cal E}) + \frac{1}{2} c_1 ({\cal L}_n)^2,
\label{shift} \end{equation} since $\it{ch}({\cal E}) \otimes
\it{ch}({\cal L}) = \it{ch}({\cal E}) \it{ch}({\cal L})$. Hence we
can effectively concentrate on the $SU(N)$ component of the
partition function, from which the full answer can be trivially
reconstructed by summing over the choice of such line bundles,
${\cal L}_n$.

Now we can use the action of the toric symmetry to considerably simplify the problem. There is a equivariant
localization theorem which relates the cohomologies of the moduli space ${\cal M}_{n, [A]}$ of bundles (or rather
sheaves) on ${\cal C}$ with Chern classes $n \in H_0 = \mathbb{Z}$ and $[A] \in H_2$ to those of invariant
sheaves,
\begin{equation} \chi \left( {\cal M}_{n, [A]} \right) = \chi \left( {\cal M}_{n, [A]}^T \right), \end{equation}
that is, the Euler character of the moduli space is captured by the considering only the equivariant bundles. In
more physical terms, the partition function involves summing over BPS states of D2 branes wrapping the legs of the
toric diagram, that is the torus invariant $\P^1$'s, and D0 branes at the equivariant points, which are the
vertices, bound to the D4 brane wrapping the nonreduced cycle ${\cal C}$.

In the next section, we will determine the number of equivariant bound states with wrapped D2 branes, and compute
the induced D0 charge. They will correspond to a choice of representation along each leg, with rank determined by
the D4 charges (for example, we will have $U(N)$ representations for D2 branes bound to a stack of $N$ D4 branes).
By cutting the Calabi Yau along the legs of the toric diagram, to effectively obtain glued $\C^3$ geometries.
Clearly this cutting of the Calabi Yau is only valid in the toric context, since we have seen that the sum over D2
branes localizes to those wrapping only the legs. The situation here is analogous to that studied in \cite{AKMV}
where the topological string amplitudes on toric Calabi Yau were shown to reduce to gluing copies of the $\C^3$
vertex, because the worldsheet instantons localized to equivariant curves. Similar localization arguments were
used by \cite{qfoam} in the more precisely related context of counting instantons of the $U(1)$ gauge theory
living on a D6 brane, and we will not repeat them here. This defines a vertex, ${\cal V}_{R Q P}$, depending on up
to three intersecting stacks of D4 branes intersecting in $\C^3$, and the asymptotic representations, $R, Q, P$,
associated to the D2 configuration. Gluing these vertices by the propagators along the legs, we obtain the full
partition function.

In the vertex geometry, the torus invariant bound states of D0
branes to the D4 worldvolume in the nilpotent Higgs phase are
given by the equivariant ideals of the algebra \eqref{surf}. The
toric symmetry acts by $(\C^*)^3$ multiplication on the
coordinates, $(x,y,z) \rightarrow (\lambda_1 x, \lambda_2 y,
\lambda_3 z)$, hence the only invariant ideals are those generated
by monomials. Depicting the monomials, $x^n y^m z^k$, of the
polynomial algebra, $\C[x,y,z]$, as the lattice of points,
$(n,m,k)$, in the positive octant, the algebra relation $x^N y^M
z^K = 0$ is represented by deleting all points (monomials) set to
zero. The remaining points give a three dimensional partition,
which has been "melted" from the corner of the cubic lattice.

Consider a D0 bound state described by a toric ideal, ${\cal I}
\lhd {\cal A}$, spanned by monomials, $x^n y^m z^k$, for $(n,m,k)
\in S$. Clearly if $x^n y^m z^k \in {\cal I}$, then so is $x^{n+1}
y^m z^k$ and so on, hence the quotient algebra $\frac{{\cal
A}}{{\cal I}}$ will be depicted as a three dimensional partition,
in the truncated region described by the D4 worldvolume algebra,
${\cal A}$. This quotient is the natural algebra associated to the
D0 worldvolume, the codimension 2 analog of the divisor of the
sheaf on ${\cal C}$. Hence the D0 charge of this bound state is
exactly the dimension of the quotient algebra, which is the number
of points in the partition. The region in which the configurations
are restricted to live is defined by the polynomial relations of
the algebra given by \eqref{surf}, shown graphically in figure 4.

\end{section}

\begin{section}{Propagators and gluing rules}

Typically, our D4 branes will wrap 4-cycles containing compact toric divisors joining two fixed point vertices.
These geometries will always look locally like ${\cal O}(-p) \rightarrow \P^1$, for some $p$, possibly
intersecting other D4 branes along the toric divisors in the Calabi Yau, $X$. For simplicity, we will first
analyze this geometry in isolation, with a single stack of $N$ wrapped D4 branes. They can be described in the
above language as the subscheme, ${\cal C} \subset X$, associated to the graded algebra \begin{equation} {\cal A}
= \frac{\C [x:t;y,z]}{z^N},
\end{equation} with weights defined by \begin{equation} (x:t;y,z) \sim (\lambda x: \lambda t; \lambda^{-p} y,
\lambda^{p-2} z ), \end{equation} for $\lambda \in \C^\times$.

Rank $N$ vector bundles on the 4-cycle can now be described as rank $1$ sheaves over ${\cal C}$, and thus by
ideals of ${\cal A}$. Moreover, invariant bundles whose corresponding equivariant divisors are (possibly non-reduced)
multi-wrappings of the $\P^1$ are equivalent to homogeneous ideals of the form
\begin{equation} {\cal I} = ( \{y^{R_i} z^{i-1} \} ),
\end{equation} for $i=1, \dots N$. These ideals are in one to one correspondence with irreducible representations
of $U(N)$ with positive row lengths. More general representations can be obtained as explained above in equation
\eqref{shift}. We proceed to calculate the Chern characters of such bundles, which will serve as the background
configuration of our crystals. This might seem difficult, since the base manifold is now non-reduced, so we will
need to relate these sheaves to ones living over the full Calabi-Yau geometry.

In a similar manner, the propagator along legs where stacks of D4
branes intersect can be determined by counting equivariant ideals
of the algebra \begin{equation} {\cal A} = \frac{\C [x:t;
y,z]}{(y^M z^N)}. \end{equation} They will be of the form
\begin{equation} {\cal I} = ( \{ y^{R_i} z^{i-1} \} ),
\label{ideal} \end{equation} where the  Young tableau $R$ lies in
the region restricted by $N$ and $M$, that is, $0 \leq R_i < M$
for $i > N$. We will refer to such Young diagrams as being of type
$(N,M)$. The following analysis of the induced D0 charge for rank
$N$ bundles is easily extended to include this case as well.

The subscheme, ${\cal C}$, is embedded via a map \begin{equation} i: {\cal C} \hookrightarrow X, \end{equation}
allowing us to push forward a sheaf, ${\cal E}$, to $i_* {\cal E}$, a sheaf on $X$ with support only along the
4-cycle. This corresponds to the fact that the ideals of ${\cal A}$ defined above are clearly also ideals of the
structure algebra of $X$ itself. The Chern characters of $i_* {\cal E}$ can be determined using equivariant
techniques to be \cite{qfoam} \begin{equation} \begin{split} 2 \it{ch}_3 = ||R^t||^2 + \frac{p}{2} \kappa_R \\
\it{ch}_2 = |R| [\mathbb{P}^1]. \label{chfoam}
\end{split} \end{equation}

This is not quite the whole answer, since those Chern characters have contributions not only from the bundle,
${\cal E}$, but from the embedding of the subscheme ${\cal C}$ itself. Therefore we must essentially subtract the
corrections arising from the normal bundle to ${\cal C}$. More precisely, the Chern character and the push forward
do not commute, in a way measured by the Grothendieck-Riemann-Roch formula \cite{SGA}, \begin{equation} \it{ch}
\left(f_* ({\cal E}) \right) = f_* \left( \it{ch} ( {\cal E}) \cdot {\mathit{Td}} ( {\cal T}_f) \right),
\end{equation} where $f: Y \rightarrow X$, $\textit{ch}$ is the Chern character, ${\mathit{ Td}}$ is the Todd
class, and ${\cal T}_f$ is the relative tangent bundle. Our
situation of an immersion into a Calabi-Yau manifold is a
significant simplification of the special case of the GRR formula
considered in \cite{Jouan}.

We find that the Chern characters are related by \begin{equation} \begin{split} \it{ch}_3(i_* {\cal E}) = i_*
\left(\it{ch}_2 ({\cal E})\right) - \frac{1}{2} &[{\cal C}] \cdot i_* \left( \it{ch}_1({\cal E}) \right) = \it{ch}_2
({\cal E}) - N \frac{p-2}{2} |R| \\
\it{ch}_2( i_* {\cal E}) &= i_* \it{ch}_1({\cal E}) = |R| \ [\P^1] , \label{ch} \end{split} \end{equation} up to
trivial terms that are independent of $R$. Applying this in reverse to the Chern characters from the three
dimensional point of view written in \eqref{chfoam}, we conclude that the brane charges are
\begin{equation} \begin{split} q_{D 0} = \frac{p}{2} \kappa_R + ||R^t||^2 + \frac{p-2}{2} N |R| &= \frac{p}{2} C_2 (R)
+ ||R^t||^2 - N|R| \\ q_{D 2} & = |R| . \end{split} \end{equation}
It will be useful later to separate these terms into a propagator
depending on the bundle, ${\cal O}(-p)$, and a piece that will be
included in the vertex amplitude calculated below. Therefore we
find the toric propagator to be \begin{equation} q^{\frac{p}{2}
C_2(R)} e^{i \theta |R|} , \end{equation} as expected from
q-deformed Yang Mills, and the remaining pieces of the second
Chern character give the corresponding factors, \begin{equation}
\exp \left\{ -g_s \left( \half ||R^t||^2 - \frac{N}{2} |R| \right)
\right\} , \label{backgr} \end{equation} in the two vertices being
glued.

These bound D2 branes also contribute to the induced D0 charge when they intersect at the vertices. In particular,
the intersection numbers give terms of the form \begin{equation} \sum_{i=1}^N R_i Q_i, \end{equation} for D2
branes bound to the same stack of $N$ D4 branes along two intersecting legs of the toric diagram. We will see in
the next section that these terms exactly cancel a correction to the vertex associated to the number of boxes
deleted by the background D2 branes, as expected.

If there are two intersecting stacks of D4 branes, the induced D0
charge can be similarly computed as \begin{equation} q_{D0} =
\frac{p}{2} ||R||^2 + \frac{2-p}{2} ||R^t||^2 + \frac{p-2}{2} N
|R| - \frac{p}{2} M |R|, \end{equation} where the D2 bound state
is of the type found in \eqref{ideal}.

\begin{section}{The crystal partition function}

We will use the transfer matrix method, described, for example, in
\cite{Okounkov}, to evaluate the partition functions of these
crystal ensembles. Recall that cutting a three dimensional
partition along diagonal planes, $y=x+t$, gives a sequence of
interlacing Young diagrams, $\{ \nu (t) \}$, indexed by $t$. In
particular, \begin{equation}\nu(t+1) \succ \nu(t),
t<0,\end{equation} and \begin{equation}\nu(t+1) \prec \nu(t), t
\geq 0,\end{equation} where two Young diagrams, $\mu$ and $\nu$,
are said to interlace, denoted by $\mu \prec \nu$, if
\begin{equation}\nu_1 \geq \mu_1 \geq \nu_2 \geq \mu_2 \geq
...\end{equation} It is now convenient to use the correspondence
between Young diagrams and states of the NS sector a free complex
fermion (or boson, via bosonization) in two dimensions, to express
the crystal partition function as a sequence of operators acting
between the states associated to the asymptotic Young diagrams.

Given a Young diagram, $\nu$, define its Frobenius coordinates as \begin{equation} a_n = \nu_n - n +\half, \ \ b_n
= \nu^t_n -n+\half, \end{equation} where the index runs from $1$ to $d$, the length of the diagonal of $\nu$. Then
the fermionic state, \begin{equation} |\nu> = \prod_{n=1}^d \psi^*_{a_n} \psi_{b_n} |0>, \end{equation} is
nontrivial and uniquely associated to $\nu$, since the $a_n$ and $b_n$ are distinct and uniquely determine $\nu$.

The key point in this transfer matrix approach is that the
operators \begin{equation} \Gamma_\pm (z) = \exp \left( \sum_{n>0}
\frac{z^{\pm n} J_{\pm n}}{n} \right),\label{gammadef}
\end{equation} constructed out of the modes, $J_n$, of the fermion
current, $\psi^* \psi$, generate the interlacing condition. That
is, \begin{equation}\Gamma_+ (1) |\nu> = \sum_{\mu \prec \nu}
|\mu>, \textrm{ and} \end{equation}\begin{equation} \Gamma_- (1)
|\nu> = \sum_{\mu \succ \nu} |\mu>.\end{equation} The number of
boxes in a single slicing is easily computed using the Virasoro
zero mode, $L_0$, to be
\begin{equation}q^{L_0} |\nu> = q^{|\nu|} |\nu>,  \end{equation} which can be applied sequentially to evaluate the
crystal action of a given partition.

It is important to take into account a finite boundary effect
arising from the difference between the number of boxes falling
inside a region bounded by diagonal slicings as opposed to a
rectangular box. If we regulate the crystal partition function by
placing it in a rectangular box of length $L$ along some axis,
then for $L$ sufficiently large, only the background asymptotic
representation, $R$, will contribute near the boundary. We will
count the total number of boxes in the three dimensional
partition, and hence must subtract off the background
contribution, $q^{L|R|}$. The diagonal slicing includes an
additional number of boxes, which are easily counted to be
\begin{equation} \half ||R||^2 - \half |R| = \half \sum_i R_i
(R_i-1), \end{equation} or its transpose, depending on the
orientation of $R$ as in \cite{topcrystal}.

This is still not quite right when there are asymptotic
representations along more then one leg, since the subtraction of
$L_1 |R| + L_2 |Q| + L_3 |P|$ over counts the number of boxes in
the background by the intersection at the origin.\footnote{This
mistake has actually appeared frequently in the literature.} For
example, the intersection of two partitions with the parallel
orientation has \begin{equation} \sum_i R_i Q_i \label{twoint}
\end{equation} points. It will turn out that this effect is cancelled by the contribution of the Chern classes of
the background divisors themselves.

Define the normalized vertex in terms of the crystal partition function, $P(R^a, N_a)$, to be, schematically,
\begin{equation} {\cal V}^{(N_a)}_{R^a} = \exp \left\{\frac{g_s}{2} \left( \sum_a N_a |R^a| - \sum_a ||R^a||^2 + 2
\sum_{a<b} R^a \cdot R^b - 4 R^1 \cdot R^2 \cdot R^2 \right)
\right\}  P(R^a, N_a), \label{norm} \end{equation} where $a$
ranges over the nontrivial asymptotics, $R^a \cdot R^b$ is given
by \eqref{twoint}, the final term is the triple intersection, and
transposes must be used appropriately depending of the orientation
of the Young tableaux. These terms are perfectly cancelled by the
Chern classes of the background bundle, which we determined in
\eqref{backgr}.

We will make extensive use of various identities involving the
operators, $\Gamma$, and skew Schur functions, which are collected
in Appendix A.

\subsection{One stack of $N$ D4 branes}

\begin{figure}
\begin{center}
\epsfig{file=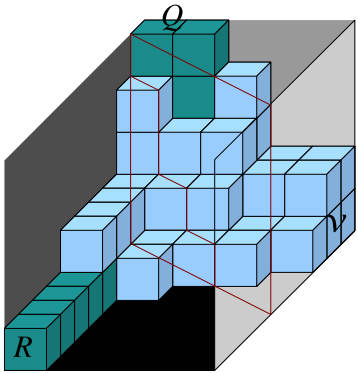,height=5cm} \caption{Shown in red is a
diagonal slice of the crystal describing 5 coincident D4 branes in
the vertex. The meaning of the representation $\nu$ will be
discussed in section 6.}\label{crystal1}
\end{center}
\end{figure}

Let us apply these methods first to the simplest case of a single stack of D4 branes. Our analysis in this
subsection will parallel that of Okuda \cite{Okuda}, where the special case of a single nontrivial asymptotic
representation was discovered in the context of Chern-Simons theory.\footnote{In fact, the original motivation of
this work was to understand the underlying structure of that relation!}

Regulating the crystal by placing it in a box of size $L \times  N
\times \infty$, we apply the transfer matrix method as shown in
figure 1, in a background determined by the representation $Q$
along the vertical leg, with asymptotic $R$ to the left and
$\cdot$ required by the truncation on the right. The pattern of
interlacing operators is altered by the presence of $Q$, as
explained in \cite{topcrystal}, thus we see that the normalized
partition function defined by \eqref{norm} is
\begin{equation} {\cal S}_{R Q} = q^{\frac{N}{2} (|R|+|Q|)}  q^{\frac{1}{2} ||Q^t||^2 + \frac{1}{2} |R|} q^{-L|R|}
<R| \prod_{n>0}^L \left(
q^{L_0} \Gamma_{\pm} (1) \right) q^{L_0} \prod_{m=1}^N \left( \Gamma_{\mp} (1) q^{L_0} \right) |0>,
\end{equation} where the pattern of pluses are minuses is determined by the shape of $Q$. Commuting the
$q^{L_0}$'s to the outside, and splitting the middle one in half, we obtain \begin{equation}q^{\frac{1}{2}
||Q^t||^2} <R| \prod_{n>0} \Gamma_\pm (q^{-n+\half}) \prod_{m=1}^N \Gamma_\mp (q^{m -\half}) |0>, \end{equation}
in the $L \rightarrow \infty$ limit. Next, we commute all of the $\Gamma_+$ to the right, where they act trivially
on $|0>$, noting that the commutator, $Z$, depends only on the representation $Q$. This gives
\begin{equation} q^{\frac{1}{2} ||Q^t||^2} <R|\prod_{m=1}^N \Gamma_- (q^{-Q_m + m - \half}) |0> Z(Q) =
q^{\frac{1}{2} ||Q^t||^2} s_{R} (q^{-Q_m+m-\half}) Z(Q).\end{equation} To determine $Z(Q)$, note that
\begin{equation}{\cal S}_{\cdot Q} = q^{\frac{1}{2} ||Q^t||^2-\frac{N}{2} |Q|} Z(Q) = {\cal S}_{Q \cdot} = Z(\cdot)
s_{Q} (q^{-\rho_N}).\end{equation} Therefore, pulling everything
together, we find that \begin{equation} {\cal S}_{R Q} = {\cal
S}_{\cdot \cdot} \ s_{Q}(q^{-\rho_N}) s_{R} (q^{-\rho_N-Q}),
\end{equation} which is exactly the Chern Simons link invariant
obtained from q-deformed Yang Mills in \cite{AOSV} and \cite{AJS}!

\begin{figure}
\begin{center}
\epsfig{file=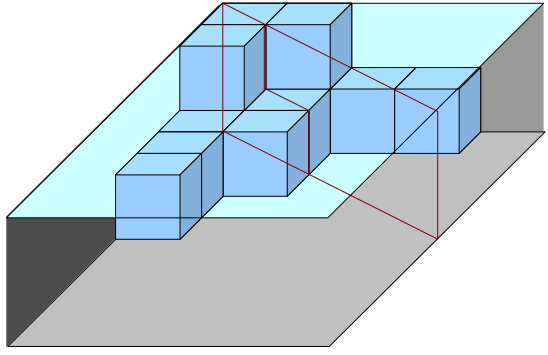,height=5cm} \caption{Here we rotate the
crystal with trivial asymptotics associated to the $U(2)$ theory
with no bound D2 branes.}\label{crystal2}
\end{center}
\end{figure}

For future reference, we will derive another useful formula for ${\cal S}_{R Q}$ by slicing the crystal with a
different orientation as depicted in figure 2. Applying manipulations similar to the above, we find that
\begin{equation}\begin{split}{\cal S}_{R Q} = q^{\half (-N |R|-N |Q| + ||R||^2+||Q||^2)}& q^{-\half ( ||R^t||^2+||Q^t||^2)}
 \\ &\times <R^t| \prod_{n>0} \Gamma_+ (q^{-n+\half}) {\cal P}^t_N \prod_{m>0} \Gamma_- (q^{m-\half}) |Q^t>, \end{split}
\end{equation} where ${\cal P}_N$ (${\cal P}^t_N$) is the projection operator onto the space of
representations with at most $N$ rows (columns). This implements
in the transfer matrix language the fact that the three
dimensional partition has height truncated by $N$, by enforcing
the constraint on the largest diagonal tableaux. Writing this in
terms of Schur functions, we find that
\begin{equation}{\cal S}_{R Q} = {\cal S}_{\cdot \cdot} \  (-)^{|R|+|Q|} q^{-\half (C_2(R)+C_2(Q) )} \sum_{P
\in U(N)} s_{P / R} (q^{\rho}) s_{P / Q} (q^{
\rho}),\label{CSform} \end{equation} where we have used
\eqref{schur} to remove the transposes.

\subsection{$N$ D4 branes intersecting $M$ D4 branes}

Suppose we have $N$ D4 branes wrapping the $x-y$ plane, $M$ in the
$y-z$ plane, with asymptotic conditions defined by a $U(N)$
representation, $R$, along the $x$-axis, and a $U(M)$
representation, $Q$, in the $z$ direction. Then we will proceed to
check that the D4 brane partition function is given by the vertex
$\mathcal{V}_{R Q}$ of the statical ensemble of three dimensional
partitions restricted to lie within the volume of the D4 branes
with asymptotics $R$ and $Q$, normalized according to \eqref{norm}
as before. Using the transfer matrix method, with diagonal
slicings oriented in the planes $y=x+n$, indexed by $n$ (see
figure 3), we obtain

\begin{equation} \begin{split} {\cal V}^{(N,M)}_{R,Q} =& q^{\half ||R||^2+\half ||Q^t||^2 - \frac{N}{2} |R| -
\frac{M}{2} |Q|} q^{-L|R|} q^{-\half ||R||^2 + \half |R|} \times \\ & <R^t| \prod_{n=1}^{L-M} \left( q^{L_0}
 \Gamma_+ (1) \right) q^{L_0} {\cal P}^t_N \prod_{m=1}^M \left( \Gamma_\pm (1) q^{L_0} \right) \prod_{k >0}
 \left( \Gamma_\mp (1) q^{L_0} \right) |0>, \end{split} \end{equation} where $Q$ determines the pattern of $\Gamma_\pm$, as
before, and the position of the projection, ${\cal P}^t_N$, is
fixed by the truncation constraint that the point $(N+1, M+1, 1)$
is not in the partition. We commute the $L_0$'s to the outside,
dividing them at the plane $y=M+z$ where we enforce the
projection, to obtain in the infinite $L$ limit, \begin{equation}
\begin{split} q^{-M |R|-\half C_2(R)+\half ||Q^t||^2 - \frac{M}{2}
|Q|} \sum_{A \in U(N)} <R^t| \prod_{n>0} \Gamma_+ (q^{-n+\half})
|A^t> & \times
\\ <A^t| \prod_{m=1}^M \Gamma_\pm (q^{m-\half}) & \prod_{k>0} \Gamma_\mp (q^{M+k-\half})
|0>. \label{twoder} \end{split} \end{equation}

\begin{figure}
\begin{center}
\epsfig{file=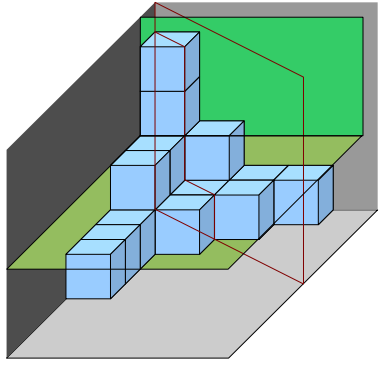,height=5cm} \caption{The crystal for $N
= 2$ intersecting $M = 1$ D4 branes is oriented as shown, where
the green planes are the boundaries of the allowed
region.}\label{crystal3}
\end{center}
\end{figure}

Now, commuting the remaining $\Gamma_+$ to act trivially on $|0>$,
we pick up a factor, $Y(Q)$, independent of $R$. Writing the
resulting expression is terms of Schur functions, we have shown
that
\begin{equation}{\cal V}_{R Q} = q^{-M |R|-\half C_2(R)+\half ||Q^t||^2 - \frac{M}{2} |Q|} Y(Q) \sum_{A \in U(N)}
s_{A^t/R^t} (q^{-\rho}) s_{A^t} (q^{-Q^t-\rho})
q^{M|A|}.\end{equation} Note that the commutator term, $Y(Q)$, can
be expressed as
\begin{equation}<0| \prod_{m=1}^M \Gamma_\pm (q^{m-\half}) \prod_{k>0} \Gamma_\mp (q^{M+k-\half}) |0>,
\end{equation} which, up to a shift in the origin of slicings by $M$ units, is exactly the $M$-truncated crystal
with asymptotics $Q$ and $\cdot$. Thus the results of the previous
subsection imply that \begin{equation}Y(Q) = q^{\frac{M}{2} |Q|}
q^{-\half ||Q^t||^2} s_{Q} (q^{\rho_M}). \end{equation}

Manipulating our expression with the Schur function identities \eqref{schur} and \eqref{schsum}, we find
\begin{equation} \begin{split} q^{-M |R|} q^{-\half C_2(R)} \sum_\mu \left( (-)^{|R|+|\mu|} \sum_{A \in U(N)}
s_{A/R} (q^{\rho}) s_{A / \mu} (q^\rho) \right) q^{\frac{M}{2}
|\mu|} \\  \times \left( s_Q (q^{\rho_M}) s_{\mu} (q^{Q+\rho_M})
\right). \label{pardec} \end{split} \end{equation}

We can now recognize the terms in parenthesis in \eqref{pardec} to be exactly the Chern-Simons link invariants,
$q^{\half C_2(R) + \half C_2 (\mu)} S^{(N)}_{R \mu} (q)$ and $S^{(M)}_{\mu Q} (q^{-1}) = S^{(M)}_{\bar{\mu} Q} $.
Therefore, recalling that $C^{(N)}_2 (P) = C_2^{(M)} (P) + (N-M)|P|$ for a $U(M)$ representation $P$, we find that
\begin{equation}{\cal V}_{R Q} = q^{-M|R|} \sum_{P \in U(M)} S^{(N)}_{R P} q^{\half C^{(M)}_2 (P)} q^{\frac{N}{2}
|P|} S^{(M)}_{\bar{P} Q},\end{equation} noting that $S^{(M)}_{P Q}$ vanishes unless $P$ is a $U(M)$
representation. This is in complete agreement, up to overall $U(1)$ factors that were not carefully taken into
account, with the result of \cite{AJS}, where this partition function was derived by reducing the D4 brane
worldvolume theories to q-deformed Yang-Mills in two dimensions, coupled by bifundamental fields living at the
intersection point.

Thus far we have only considered the case of trivial asymptotics along the intersection plane of the branes, which
is natural assuming that direction is noncompact in the toric Calabi-Yau. Lifting that restriction, we proceed
much as before after replacing $|0>$ by $|P^t>$ in \eqref{twoder}, to obtain \begin{equation} \begin{split}
q^{-M|R|-\half ||R||^2 -\half ||P||^2+\frac{M}{2} |Q|-\half ||Q^t||^2} (-)^{|R|} \times & \\ \sum_{\mu,
\nu} \left(\sum_{A \in U(N)} s_{A / R} (q^{\rho}) s_{A / \nu} (q^\rho) \right) \ q^{\frac{M}{2} |\nu|} s_Q
(q^{\rho_M}) & s_{\nu / \mu} (q^{Q+\rho_M}) (-)^{|\mu|} s_{P^t / \mu^t} (q^{\bar{Q}+\rho_M}) .\end{split}
\end{equation}
By definition of the skew Schur functions and the Chern Simons invariants, \\
 $s_Q (q^{\rho_M}) s_{\nu / \mu} (q^{Q+\rho_M)}) = N^{\nu}_{B \mu} S^{(M)}_{B Q} (q^{-1}) = N^\nu_{B \mu}
S^{(M)}_{B \bar{Q}} $, giving the final result
\begin{equation} \frac{q^{-M|R|}}{S_{\bar{Q} \cdot}} \sum_{D, B, C} S^{(N)}_{R D} q^{\half C^{(N)}_2 (D)}
q^{\frac{M}{2} |D|} \hat{N}^{D P}_{B C^t} S^{(M)}_{B \bar{Q}} S^{(M)}_{C Q}, \end{equation} where
$D$ is a representation of $U(N)$, $B$ and $C$ are representations of $U(M)$, and \begin{equation} \hat{N}^{D P}_{B C^t}
= (-)^{|D|} \sum_\mu N^D_{B \mu} (-)^{|\mu|} N^P_{C^t \mu}. \label{tensordef} \end{equation}

\subsection{The general vertex}

The most general possibility for intersecting toric D4 branes is the triple intersection with representations
running along all three legs of the vertex. The partition function is significantly more complicated, and,
although it can still be computed schematically with the transfer matrix technology, we will not be able to
express it in terms of the Chern-Simons link invariants. In the context of OSV, the existence of a large $N$
factorization remains clear, simply from the crystal picture itself, as explained in detail in the next section.
We will also consider a special case that can be solved more explicitly.

\begin{figure}
\begin{center}
\epsfig{file=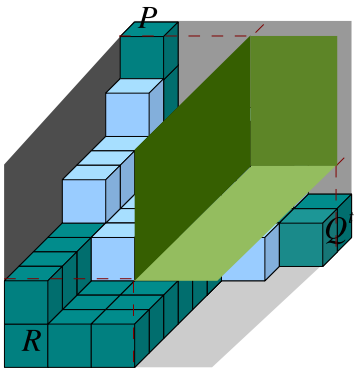,height=5cm} \caption{The restricted
region of general vertex describing $U(N) \times U(M) \times
U(K)$, for $N=3$, $M=1$, and $K=2$, in the local $\C^3$ geometry
is demarcated by the green planes, and the asymptotes are
labelled.}\label{crystal4}
\end{center}
\end{figure}

The partition function for the crystal oriented as in figure 4, but with $M \geq N$, can be determined
using the same methods to be given by
\begin{equation} \begin{split} q^{-\half ||R^t||^2 - \half ||Q||^2} & q^{(N-M)(|R|-|Q|)}  \times \\ & <R| \prod_{n>0}
\Gamma_\pm (q^{-n+\half}) {\cal P} \prod_{m=1}^{M-N} \Gamma_\pm (q^{m-\half}) \prod_{l>0} \Gamma_\mp (q^{M-N+l-\half}) |Q^t>,
\end{split} \end{equation} where we have chosen the framing with parallel rectangular walls for convenience. The pattern
of signs of $\Gamma$ is given by $P$, and the projection ${\cal P}$ forces the
Young diagram to live in the restricted region determined by $K$ in the vertical direction, and, in the horizontal
direction, by the distance along the diagonal from the background representation, $P$, to the point $(N,M)$. We
orient the representations along the intersecting pairs of D4 branes such that $R$ is of type $(N,K)$, $Q$ is
$(K,M)$, and $P$ is $(M,N)$ in the notation of \eqref{ideal}.

\begin{figure}
\begin{center}
\epsfig{file=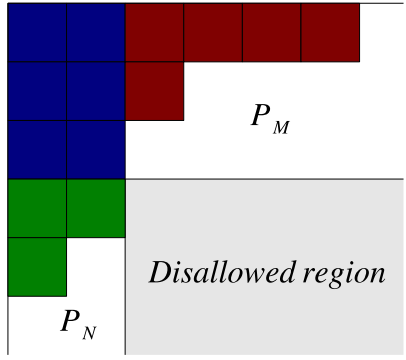,height=5cm} \caption{A special kind of $(M,N)$ Young diagram is which the $U(N)$ and
$U(M)$ degrees of freedom are decoupled is shown.}\label{decoupled}
\end{center}
\end{figure}

Let us consider the special case that all of the boxes in the bifundamental region of $P$ are filled, and joined
to a $U(N)$ representation, $P_N$, and a $U(M)$ representation, $P_M$, as depicted in figure 5, for which a more
detailed analysis is possible. This form of $P$ implies that ${\cal P} = {\cal P}^t_K$, and commuting the
$\Gamma_\pm$'s appropriately, it is convenient to define
\begin{equation} W = q^{-\half ||R^t||^2 - \half ||Q||^2} q^{(N-M)(|R|-|Q|)} X (P_N) X (P_M), \end{equation}
in terms of the commutator term $X(P_N) = q^{-\half ||P_N^t||^2+ \frac{N}{2} |P_N|} s_{P_N} (q^{-\rho_N})$, and
the framing prefactor. Then we find that the statistical partition function is
\begin{equation} W \sum_{L \in U(K)} \sum_{\mu, \nu} s_{R / \mu} (q^{-\frac{N}{2}+\rho_N+\bar{P}_N}) s_{L^t/ \mu}
(q^{-\rho+N-P_N^t}) s_{L^t / \nu} (q^{M -\rho -P_M^t}) s_{Q^t / \nu} (q^{-\frac{M}{2} +\rho_M+ {\bar P}_M}).
\end{equation} Now apply various identities to expand and simplify Schur functions evaluated at expressions of the form
$q^{\rho + P_N^t}$, to obtain,
\begin{equation} \begin{split} W \sum_{L \in U(K)} \sum_{\mu, \nu, \alpha, \beta}
q^{-\frac{N}{2} |R|-\frac{N}{2} |\mu| + (M+N) |L| -\frac{M}{2} |\nu| -\frac{M}{2} |Q|} (-)^{|\mu|+|\nu|}  s_{R
/ \mu} (q^{\rho_N+\bar{P_N}}) \ \times
\\ s_{\alpha / \mu^t} (q^{-\frac{N}{2}+\rho_N+P_N}) s_{L/\alpha}(q^{\rho-N}) s_{L/\beta}(q^{\rho-M}) s_{\beta /
\nu^t}(q^{-\frac{M}{2}+\rho_M+P_M}) s_{Q^t / \nu}(q^{\rho_M+\bar{P_M}}). \end{split} \end{equation} Recognizing
the appearance of the Chern-Simons link invariant from \eqref{CSform}, we can simplify the partition function to
find
\begin{equation} \begin{split} W \sum_{A,B \in U(K)} \sum_{\mu, \nu} q^{-\frac{N}{2} |R| + \frac{N}{2} |A|+ \frac{M}{2}|B|
- \frac{M}{2} |Q|} (-)^{|\mu|+|\nu|+|A|+|B|}  s_{R / \mu}
(q^{\rho_N+\bar{P_N}}) \ \times
\\ s_{A / \mu^t} (q^{\rho_N+P_N}) S^{(K)}_{ A B} q^{\half C_2(A)+ \half C_2 (B)} s_{B / \nu^t}(q^{\rho_M+P_M})
s_{Q^t / \nu}(q^{\rho_M+\bar{P_M}}) \end{split} \end{equation} Therefore putting everything together, and
rewriting the result in terms of Chern-Simons invariants, we conclude that,
\begin{equation} \begin{split}
Z = q^{(\frac{N}{2}-M) |R| + (\frac{M}{2}-N) |Q| - \half||R^t||^2 - \half ||Q||^2} q^{-\half||P_N^t||^2-\half||P_M^t||^2+
\frac{N}{2} |P_N| +\frac{M}{2} |P_M|} \sum_{A,B \in U(K)} \\ \sum_{C,D \in U(N)} \sum_{E,F \in U(M)} \hat{N}^{A R^t}_{C D^t}
 \hat{N}^{B Q}_{E F^t} q^{\frac{N}{2} |A| +\frac{M}{2}|B| + \half C_2(A) + \half C_2(B)} S^{(N)}_{C \bar{P_N}}
 S^{(N)}_{D P_N} S^{(M)}_{E \bar{P_M}} S^{(M)}_{F P_M},
\end{split} \end{equation} where the fusion coefficients are defined by equation \eqref{tensordef}.

\end{section}

\end{section}

\begin{section}{Chiral factorization and conjugation involution}

The chiral factorization of the indexed entropy at large D4 charge can be seen very elegantly in the melting
crystals. The appearance of one factor of the topological string is immediate, since in the limit that $N$ goes to
infinity, the restrictions on the tableaux become vacuous, and we exactly reproduce the propagators and crystal
vertices of the topological A-model in the same toric geometry. Referring to the derivation of the induced brane
charges \eqref{ch}, we see that one obtains exactly the framing factors and vertex amplitudes of the topological
A-model on the Calabi Yau \cite{qfoam}. Moreover, the extra contribution to the induced D0 charge is precisely
\begin{equation} \frac{1}{2} [{\cal C}] \cdot [D],\end{equation} the intersection number of the D4 branes on $[{\cal C}]$
with the homology class, $[D]$, wrapped by the D2 branes. This
combines with the chemical potential for the D2 charge to give the
attractor value of the Kahler moduli, \begin{equation} t_a =
\frac{1}{2} g_{top} [{\cal C}] \cdot [D_a] - i \theta_a,
\end{equation} for the curve $[D_a] \in H_2 (X)$.

To see what the anti-chiral block arises from, recall that the propagator depends on the quadratic Casimir, so we
expect both small representations and their conjugates to contribute, following \begin{equation} C_2({\cal
R})=\kappa_{R+} + \kappa_{R-} + N(|R_+|+|R_-|)+ N{\ell_R}^2 + 2 {{\ell}_R}(|R_+|-|R_-|), \end{equation} where
${\cal R} = (R \bar{Q})[\ell_R]$ is the $U(N)$ representation with chiral part $R_+$, anti-chiral component $R_-$,
and overall $U(1)$ factor $\ell_R$.

There is an involution on restricted three dimensional partitions of the type we have been discussing which is
inherited from the conjugation of $U(N)$ representations. Consider two successive Young tableaux in the transfer
matrix approach, $R \succ Q$, that is $R_1 \geq Q_1 \geq R_2 \geq Q_2 \geq \dots$. Then, when both are $U(N)$
representations, under conjugation, one can see that $\bar{Q}_N \geq \bar{R}_N \geq \dots$, and thus $\bar{Q}
\succ \bar{R}$, up to a shift by a sufficiently large and positive $U(1)$ factor. Therefore this defines an
involution on the chiral fermion Hilbert space, sending \begin{equation} |R> \rightarrow |\bar{R}> , \ q^{L_0}
\rightarrow q^{-L_0} \ \text{and} \ \Gamma_- \rightarrow \Gamma_+ , \end{equation} since $\Gamma$ generates the
interlacing relation, and $|\bar{R}| = - |R|$. Naturally, these relations make sense only when we are restricted
to $U(N)$ representations.

The involution also acts naturally when the slicing is oriented differently, for example, suppose $R$ and $Q$ are
$U(N)$ representations such that $R^t \succ Q^t$. It is easy to see that $(\bar{R})^t_n + (R)^t_{A-n} = N,$ for
conjugation with shifting by some sufficiently large $U(1)$ factor $A$. Thus $(\bar R)^t \prec (\bar Q)^t$, again
up to an appropriate $U(1)$ shift, chosen so that all column lengths are positive.

Applying this transformation to the interlacing representations of the crystals, one can show that they are
invariant. For a single stack of D4 branes, for example, we have \begin{equation} \begin{split} S^{(N)}_{\bar{R}
\bar{Q}} & = q^{-\half C_2 (\bar R)-\half C_2 (\bar Q)} <(\bar R)^t| \prod_{n>0} \Gamma_+ (q^{-n+\half}) {\cal
P}^t_N \prod_{m>0} \Gamma_- (q^{m-\half}) |(\bar Q)^t> \\
& = q^{-\half C_2 (R)-\half C_2 (Q)} <R^t| \prod_{n>0} \Gamma_-(q^{n-\half}) {\cal P}^t_N \prod_{m>0} \Gamma_+
(q^{-m+\half}) |Q^t> = S^{(N)}_{R Q}, \end{split} \end{equation} by commuting the $\Gamma$'s in the obvious way,
using crucially the fact that we are free to shift by $U(1)$ factors to avoid having rows of negative length.

Therefore the factorized form $Z_{YM} \sim |\psi_{top}|^2$,
follows automatically, by breaking all of the representations
associated to D2 branes wrapping the legs into their chiral and
anti-chiral pieces. The sum over chiral blocks (ie. ghost branes)
first found in \cite{AOSV} arises because crystals with order $N
|A|$ boxes can connect the chiral and anti-chiral regions with the
ghost Young tableau $A$ via a $N \times A$ transversal
configuration, contributing $e^{-g_s N |A|}$ to the partition
function, which remains finite even at large $N$. The simplest
example of this is the vertex partition function of $N$ coincident
D4 branes, where in the large $N$ limit diagrams, $\nu$, along the
truncated direction, as shown in figure 1, are exactly the ghost
tableaux, which clearly have no restrictions on their rank. The
additional D0 branes is the corner contributing to the chiral
block will experience an effective background described by the
topological vertex $C_{Q_+^t R_+ \nu}$, as expected from the
factorization formulae of \cite{AOSV}, \cite{ANV}, and \cite{AJS}.

\end{section}

\begin{section}{Concluding remarks and further directions}

We have solved the theory of BPS bound states of D2 and D0 branes
to D4 branes wrapping deformable cycles in toric Calabi-Yau by
localization, obtaining a truncated crystal partition. This
calculation is exact, even for small charges, and smoothly
asymptotes, in the OSV large charge limit, to the square of the
topological string amplitude in the crystal vertex expansion. The
crystal configurations we have found count the extra D0 brane
bound states at the singular points of the non-reduced surface
obtained by giving nilpotent VEVs to the worldvolume adjoint
fields controlling the normal deformations of the D4 branes in the
Calabi Yau. These are the toric invariant, zero size limit of the
extra 2-cycles in $H_2({\cal C}) > H_2(X)$, in the phase where the
D4 charge results from a single brane wrapping the high degree,
smooth surface ${\cal C}$.

The index of BPS bound states should be the same in all three
cases, since it is protected by supersymmetry. We check that the
results found here agree with the topological gauge theory
calculations in known examples. This gives a nice realization of
the underlying geometric nature of the interacting topological
Yang Mills theories, since the crystal picture is not at all
manifest in the quiver of four dimensional $U(N_i)$ gauge theories
with bifundamental couplings along the intersections.

We view this work as a step in the development of a nonperturbative topological vertex, that is a general method
to compute the partition function of D4 branes on toric Calabi Yau manifolds in the OSV ensemble, which defines
the nonperturbative completion of the topological A-model amplitudes. We have computed the relevant vertex and
gluing rules, whose analogs in the topological string were sufficient to solve the theory, since the A-model only
involves holomorphic maps from curves to the Calabi Yau.

The main problem in the exact OSV calculation on toric Calabi Yau
which remains to be solved is the effect of branes wrapping
compact 4-cycles. In toric Calabi Yau, compact 4-cycles always
have positive curvature, and therefore negative self-intersection.
As emphasized in \cite{AJS}, we must wrap D4 branes on very ample
divisors to create black holes with large positive values for the
geometrical moduli at the attractor point. This is the regime of
validity of the OSV conjecture. Moreover, the theory living on D4
branes instead wrapping rigid divisors with negative
self-intersection cannot be solved by the ${\cal N}=1$ mass
deformation of \cite{VW}, and it exhibits background moduli
dependance and lines of marginal stability where the number of BPS
states jump.

This might seem to preclude the analysis of any more complicated
geometries with multiple compact 4-cycles in the toric context,
such as the resolution of $A_3$ ALE fibred over $\P^1$. However,
if the D4 branes wrap toric 4-cycles, including some rigid cycles,
which still sum to a very ample divisor, or equivalently, give
rise to positive Kahler moduli at the attractor point, then there
is no reason to expect a breakdown of the OSV conjecture.
Therefore it seems likely that the effect of intersection with
sufficiently many noncompact branes wrapping "good" 4-cycles will
eliminate the problems of the twisted ${\cal N}=4$ theory on
compact rigid surfaces. It would be very interesting to see how
this works in detail, and to understand the new contributions
arising from such branes. This could lead to a more complicated
set of gluing rules, involving additional contributions from
"loops" in the toric diagram as well as vertex and propagator
terms that we have found here.

It would also be interesting to investigate the relation between
our crystals and those studied in \cite{Halm}, which have a
different kind of truncation, arising purely in the context of the
$q$ expansion of (perturbative) topological string theory. This
might further elucidate why the Chern Simons amplitudes
$S^{(N)}_{R Q}$ appear, even at finite $N$, in the solution of D4
brane $U(N)$ gauge theories.

\end{section}

\section*{Acknowledgements}

I would like to thank Mina Agaganic, Joe Marsano, Natalia Saulina,
Xi Yin, and especially Cumrun Vafa for valuable discussions. This
work was supported in part by NSF grants PHY-0244821 and
DMS-0244464. I would like to thank the Simons Workshop at the Yang
Institute of Theoretical Physics at Stonybrook, where part of this
work was conducted, for hospitality.

\begin{section}* {Appendix A}

All $\Gamma_+$ commute, as do $\Gamma_-$, however one can show
that \begin{equation} \Gamma_- (x) \Gamma_+ (y) = \Gamma_+ (y)
\Gamma_- (x) \left(1- \frac{y}{x} \right). \end{equation} Also,
\begin{equation} \Gamma_\pm (x) q^{n L_0} = q^{n L_0} \Gamma_\pm
(x q^{-n}). \end{equation} These operators are conjugate;
following the definition \eqref{gammadef} it is easy to see that
\begin{equation} \left( \Gamma_+ (x) \right)^\dag = \Gamma_-
(x^{-1}) .\end{equation} The repeated application of the
$\Gamma_\pm$ operators results in the appearance of skew Schur
functions, since
\begin{equation} \prod_n \Gamma_- (x_n) |\nu> = \sum_{\mu \supset
\nu } s_{\mu / \nu} (x_n) |\mu>.
\end{equation} We will often come across skew Schur functions
evaluated at special values of the form $x_n = \rho_n + \nu_n$,
where $\rho = \{-\half, -\frac{3}{2}, \dots \}$, and $\nu$ is a
Young diagram, as well as the finite $N$ version, $x_i = \rho^N_i
+ R_i, \ i= 1, \dots N$, where $R$ is a $U(N)$ representation, and
$\rho^N_i = \frac{N+1}{2} - i$ is the Weyl vector. These functions
obey many beautiful relations, including
\begin{equation} \begin{split} s_{\mu / \nu} (q^{\rho+\eta}) &= (-)^{|\mu|+|\nu|} s_{\mu^t / \nu^t} (q^{-\rho -\eta^t}) \\
s_{\mu / \nu} (c \{x_n\} ) &= c^{|\mu|-|\nu|} s_{\mu / \nu}(x) \\
s_{\mu / \nu} (x) &= \sum_\eta N^\mu_{\nu \eta} s_\eta (x), \label{schur} \end{split} \end{equation} where
$N^\mu_{\nu \eta}$ are the tensor product coefficients. There are various useful summation formulae as well,
\begin{equation}
\begin{split} \sum_\eta s_{\mu/\eta} (x) s_{\nu/\eta} (y) &= \prod_{n,m} \left(1-x_n y_m \right) \sum_\lambda
s_{\lambda/\mu}(y) s_{\lambda/\nu}(x) \\
\sum_\nu s_{\mu / \nu}(x) s_\nu (y) &= s_\nu (x,y), \label{schsum} \end{split} \end{equation} where $(x,y)$
denotes the conjunction of the strings of variables $x$ and $y$.

We also make use of the quadratic Casimirs, \begin{equation}
\kappa_\mu = ||\mu||^2 - ||\mu^t||^2, \ \text{ and } \ C_2^{(N)}
(R) = N|R| + \kappa_R, \end{equation} for any Young diagram $\mu$
and $U(N)$ representation $R$.

\end{section}

 \end{document}